\documentclass[aps,prl,superscriptaddress,twocolumn,preprintnumbers,showpacs]{revtex4}

\usepackage{graphicx}
\usepackage{dcolumn}
\usepackage{bm}

\newcommand{\etal}{{\em et al.}}
\newcommand{\Vub}{$V_{ub}$}
\newcommand{\ppbar}{$\bar{p}p$}
\newcommand{\Bd}{$B^{0}$}

\newcommand{\Bu}{$B^{+}$}
\newcommand{\Bs}{$B_{s}^{0}$}
\newcommand{\Bdpipi}{\Bd $\rightarrow\pi^{+}\pi^{-}$}
\newcommand{\BdKpi}{\Bd  $\rightarrow K^{+}\pi^{-}$}

\newcommand{\BdKK}{\Bd  $\rightarrow K^{+}K^{-}$}
\newcommand{\BsKK}{\Bs  $\rightarrow K^{+}K^{-}$}
\newcommand{\BsKpi}{\Bs  $\rightarrow K^{-}\pi^{+}$}
\newcommand{\Bspipi}{\Bs $\rightarrow\pi^{+}\pi^{-}$}
\newcommand{\BR}{\ensuremath{\mathcal B}}

\newcommand{\Bhhgen}{$B^{0}_{(s)}\rightarrow h^{+}h'^{-}$}
\newcommand{\dedx}{\ensuremath{\mathit{dE/dx}}}

\newcommand{\bd}{\ensuremath{B^{0}}}
\newcommand{\bs}{\ensuremath{B^{0}_{s}}}

\newcommand{\bdpipi}{\ensuremath{\bd \rightarrow \pi^+ \pi^-}}

\newcommand{\bdkpi}{\ensuremath{\bd \rightarrow K^+ \pi^-}}

\newcommand{\bskpi}{\ensuremath{\bs \rightarrow K^- \pi^+}}

\newcommand{\bskk}{\ensuremath{\bs \rightarrow  K^+ K^-}}
\newcommand{\bspipi}{\ensuremath{\bs \rightarrow  \pi^+ \pi^-}}
\newcommand{\bdkk}{\ensuremath{\bd \rightarrow  K^+ K^-}}

\newcommand{\fracpipisukpi}{\ensuremath{\frac{\mathcal{B}(\bdpipi)}{\mathcal{B}(\bdkpi)}}}
\newcommand{\frackksukpi}{\ensuremath{\frac{\mathit{f_s}}{\mathit{f_d}}\frac{\mathcal{B}(\bskk)}{\mathcal{B}(\bdkpi)}}}

\newcommand{\fraclimbskpi}{\ensuremath{\frac{\mathit{f_s}}{\mathit{f_d}}\frac{\mathcal{B}(\bskpi)}{\mathcal{B}(\bdkpi)}}}
\newcommand{\fraclimbspipi}{\ensuremath{\frac{\mathcal{B}(\bspipi)}{\mathcal{B}(\bskk)}}}
\newcommand{\fraclimbdkk}{\ensuremath{\frac{\mathcal{B}(\bdkk)}{\mathcal{B}(\bdkpi)}}}

\newcommand{\cdf}{CDF Collaboration}
\newcommand{\babar}{BaBar Collaboration}
\newcommand{\cleo}{CLEO Collaboration}
\newcommand{\belle}{Belle Collaboration}
\newcommand{\alep}{ALEPH Collaboration}

\begin{document}

\title{Observation of \BsKK\ and Measurements of Branching Fractions
of Charmless Two-body Decays of \Bd\ and \Bs\ Mesons in \ppbar\  
Collisions at $\sqrt{s}=1.96$ TeV}
\affiliation{Institute of Physics, Academia Sinica, Taipei, Taiwan 11529, Republic of China} 
\affiliation{Argonne National Laboratory, Argonne, Illinois 60439} 
\affiliation{Institut de Fisica d'Altes Energies, Universitat Autonoma de Barcelona, E-08193, Bellaterra (Barcelona), Spain} 
\affiliation{Baylor University, Waco, Texas  76798} 
\affiliation{Istituto Nazionale di Fisica Nucleare, University of Bologna, I-40127 Bologna, Italy} 
\affiliation{Brandeis University, Waltham, Massachusetts 02254} 
\affiliation{University of California, Davis, Davis, California  95616} 
\affiliation{University of California, Los Angeles, Los Angeles, California  90024} 
\affiliation{University of California, San Diego, La Jolla, California  92093} 
\affiliation{University of California, Santa Barbara, Santa Barbara, California 93106} 
\affiliation{Instituto de Fisica de Cantabria, CSIC-University of Cantabria, 39005 Santander, Spain} 
\affiliation{Carnegie Mellon University, Pittsburgh, PA  15213} 
\affiliation{Enrico Fermi Institute, University of Chicago, Chicago, Illinois 60637} 
\affiliation{Joint Institute for Nuclear Research, RU-141980 Dubna, Russia} 
\affiliation{Duke University, Durham, North Carolina  27708} 
\affiliation{Fermi National Accelerator Laboratory, Batavia, Illinois 60510} 
\affiliation{University of Florida, Gainesville, Florida  32611} 
\affiliation{Laboratori Nazionali di Frascati, Istituto Nazionale di Fisica Nucleare, I-00044 Frascati, Italy} 
\affiliation{University of Geneva, CH-1211 Geneva 4, Switzerland} 
\affiliation{Glasgow University, Glasgow G12 8QQ, United Kingdom} 
\affiliation{Harvard University, Cambridge, Massachusetts 02138} 
\affiliation{Division of High Energy Physics, Department of Physics, University of Helsinki and Helsinki Institute of Physics, FIN-00014, Helsinki, Finland} 
\affiliation{University of Illinois, Urbana, Illinois 61801} 
\affiliation{The Johns Hopkins University, Baltimore, Maryland 21218} 
\affiliation{Institut f\"{u}r Experimentelle Kernphysik, Universit\"{a}t Karlsruhe, 76128 Karlsruhe, Germany} 
\affiliation{High Energy Accelerator Research Organization (KEK), Tsukuba, Ibaraki 305, Japan} 
\affiliation{Center for High Energy Physics: Kyungpook National University, Taegu 702-701, Korea; Seoul National University, Seoul 151-742, Korea; and SungKyunKwan University, Suwon 440-746, Korea} 
\affiliation{Ernest Orlando Lawrence Berkeley National Laboratory, Berkeley, California 94720} 
\affiliation{University of Liverpool, Liverpool L69 7ZE, United Kingdom} 
\affiliation{University College London, London WC1E 6BT, United Kingdom} 
\affiliation{Centro de Investigaciones Energeticas Medioambientales y Tecnologicas, E-28040 Madrid, Spain} 
\affiliation{Massachusetts Institute of Technology, Cambridge, Massachusetts  02139} 
\affiliation{Institute of Particle Physics: McGill University, Montr\'{e}al, Canada H3A~2T8; and University of Toronto, Toronto, Canada M5S~1A7} 
\affiliation{University of Michigan, Ann Arbor, Michigan 48109} 
\affiliation{Michigan State University, East Lansing, Michigan  48824} 
\affiliation{Institution for Theoretical and Experimental Physics, ITEP, Moscow 117259, Russia} 
\affiliation{University of New Mexico, Albuquerque, New Mexico 87131} 
\affiliation{Northwestern University, Evanston, Illinois  60208} 
\affiliation{The Ohio State University, Columbus, Ohio  43210} 
\affiliation{Okayama University, Okayama 700-8530, Japan} 
\affiliation{Osaka City University, Osaka 588, Japan} 
\affiliation{University of Oxford, Oxford OX1 3RH, United Kingdom} 
\affiliation{University of Padova, Istituto Nazionale di Fisica Nucleare, Sezione di Padova-Trento, I-35131 Padova, Italy} 
\affiliation{LPNHE, Universite Pierre et Marie Curie/IN2P3-CNRS, UMR7585, Paris, F-75252 France} 
\affiliation{University of Pennsylvania, Philadelphia, Pennsylvania 19104} 
\affiliation{Istituto Nazionale di Fisica Nucleare Pisa, Universities of Pisa, Siena and Scuola Normale Superiore, I-56127 Pisa, Italy} 
\affiliation{University of Pittsburgh, Pittsburgh, Pennsylvania 15260} 
\affiliation{Purdue University, West Lafayette, Indiana 47907} 
\affiliation{University of Rochester, Rochester, New York 14627} 
\affiliation{The Rockefeller University, New York, New York 10021} 
\affiliation{Istituto Nazionale di Fisica Nucleare, Sezione di Roma 1, University of Rome ``La Sapienza," I-00185 Roma, Italy} 
\affiliation{Rutgers University, Piscataway, New Jersey 08855} 
\affiliation{Texas A\&M University, College Station, Texas 77843} 
\affiliation{Istituto Nazionale di Fisica Nucleare, University of Trieste/\ Udine, Italy} 
\affiliation{University of Tsukuba, Tsukuba, Ibaraki 305, Japan} 
\affiliation{Tufts University, Medford, Massachusetts 02155} 
\affiliation{Waseda University, Tokyo 169, Japan} 
\affiliation{Wayne State University, Detroit, Michigan  48201} 
\affiliation{University of Wisconsin, Madison, Wisconsin 53706} 
\affiliation{Yale University, New Haven, Connecticut 06520} 
\author{A.~Abulencia}
\affiliation{University of Illinois, Urbana, Illinois 61801}
\author{D.~Acosta}
\affiliation{University of Florida, Gainesville, Florida  32611}
\author{J.~Adelman}
\affiliation{Enrico Fermi Institute, University of Chicago, Chicago, Illinois 60637}
\author{T.~Affolder}
\affiliation{University of California, Santa Barbara, Santa Barbara, California 93106}
\author{T.~Akimoto}
\affiliation{University of Tsukuba, Tsukuba, Ibaraki 305, Japan}
\author{M.G.~Albrow}
\affiliation{Fermi National Accelerator Laboratory, Batavia, Illinois 60510}
\author{D.~Ambrose}
\affiliation{Fermi National Accelerator Laboratory, Batavia, Illinois 60510}
\author{S.~Amerio}
\affiliation{University of Padova, Istituto Nazionale di Fisica Nucleare, Sezione di Padova-Trento, I-35131 Padova, Italy}
\author{D.~Amidei}
\affiliation{University of Michigan, Ann Arbor, Michigan 48109}
\author{A.~Anastassov}
\affiliation{Rutgers University, Piscataway, New Jersey 08855}
% ...
\collaboration{CDF Collaboration}
\noaffiliation

\date{\today}

\begin{abstract}
We search for decays of the type \Bhhgen\ 
(where $h,h'= K$ or $\pi$) in 180~pb$^{-1}$ of \ppbar\ collisions collected
at the Tevatron by the upgraded Collider Detector at Fermilab.
We report the first observation of the new mode \BsKK\ with a yield of 
$236\pm 32$ events, corresponding to
$(f_s/f_d)\times {\mathcal B}(B^0_s\rightarrow K^+K^-)/{\mathcal 
B}(B^0\rightarrow K^+\pi^-) = 0.46 \pm 0.08(\mathit{stat.})\pm 
0.07(\mathit{syst.})$, where $f_{s}/f_{d}$ is 
the ratio of production fractions of \Bs\ and \Bd .
We find results in agreement with world averages for the \Bd\ 
modes, and set the following new limits at 90\% CL: \BR (\BsKpi) $< 5.6\times 10^{-6}$ 
and \BR (\Bspipi) $< 1.7\times 10^{-6}$.
\end{abstract}


\maketitle



The decay modes of $B$ mesons into pairs of charmless 
pseudoscalar mesons are effective probes of the quark-mixing 
(Cabibbo-Kobayashi-Maskawa, CKM) matrix
and are sensitive to potential new physics effects. Their branching
fractions and CP asymmetries can be predicted with good
accuracy and compared to rich experimental data available for \Bu\ and \Bd\ mesons, 
produced in large quantities in $\Upsilon(4S)$ 
decays~\cite{charmless-exp}. 
Measurements of similar modes predicted, but not yet observed, for the \Bs\ meson
are important to complete our understanding of $B$ meson decays. 
The measurement of observables 
from both strange and non-strange $B$ mesons allows a cancellation of 
hadronic uncertainties, thus enhancing the precision of the extraction of physics
parameters from experimental data~\cite{strangeB,Gronau:2000md,Burasetal,Bs-QCDF}. 

The branching fraction of the \BsKK\ mode is a candidate for observing an unusually large breaking of U--spin 
symmetry, and is sensitive to anomalous electroweak penguin contributions from 
new physics~\cite{Burasetal,Khodjamirian,Descotes-Genon:2006wc}. 
A combination of \Bdpipi\ and \BsKK\ observables 
has been proposed as a way to directly determine the phase of
the \Vub\ element of 
the CKM matrix (angle $\gamma$), 
or alternatively as a test of 
our understanding of dynamics of $B$ hadron decays, when compared with other 
determinations of $\gamma$~\cite{Fleischer}. The \BsKpi\ mode can 
also be used in measuring 
$\gamma$~\cite{Gronau:2000md}, and its 
CP asymmetry is a powerful model-independent test~\cite{Lipkin-BsKpi} 
of the source of the direct 
CP asymmetry observed in the \BdKpi\ 
mode~\cite{ACP-BaBelle}.
The \Bspipi\ mode proceeds only through annihilation diagrams, which 
are currently poorly known and a source of significant uncertainty in 
many theoretical calculations~\cite{B-N}. Its features are 
similar to the as yet unobserved \BdKK\ mode, but it has 
a larger predicted branching fraction~\cite{B-N,Bspipi}; a 
measurement of both modes would allow a determination of
the strength of penguin-annihilation~\cite{Burasetal}.

In this letter we report the first observation of the decay \BsKK\ and perform 
the first measurement in hadron collisions of partial widths of $B^{0}_{(s)}$ decays
to pairs of charged pions and kaons.
Throughout this paper, C-conjugate modes 
are implied and branching fractions indicate 
CP-averages unless otherwise stated.


The measurements have been performed in a sample of 180~pb$^{-1}$ of 
\ppbar\ collisions at $\sqrt{s}$ = 1.96 TeV, recorded at the 
Tevatron collider by the 
upgraded Collider Detector at Fermilab (CDF II). 
CDF II is a multipurpose magnetic spectrometer surrounded by
calorimeters and muon detectors~\cite{CDF}. The components of the detector pertinent
to this analysis are described briefly below.  
A silicon microstrip detector (SVX II)~\cite{SVX} and a cylindrical drift chamber
(COT)~\cite{COT} immersed in a 1.4 T solenoidal magnetic field
allow reconstruction of charged particles in the pseudorapidity range 
$\mid\eta\mid < 1.0$~\cite{CDF-coordinates}. 
The SVX~II consists of five concentric layers
of double-sided silicon detectors with radii between 2.5 and 10.6 cm,
each providing a measurement with 15~$\mu$m resolution in the 
$\phi$ direction.
The COT has
96 measurement layers, between 40 and 137 cm in radius, organized into 
alternating axial and $\pm 2^{\circ}$ stereo superlayers.
The transverse momentum resolution is $\sigma_{p_{T}}/p_{T} \simeq 
0.15\%\, p_{T}$/(GeV/$c$). 
The specific energy loss (\dedx) of charged particles in the COT can
be measured from the collected charge,
which is encoded in the output pulse-width of each wire.


Data were collected by a three-level trigger system, 
using a set of requirements dedicated to $B$ hadron decays into 
charged particle pairs.
At Level~1, charged particle tracks are reconstructed in the COT 
transverse plane by a hardware
processor (XFT)~\cite{xft}. Two opposite-curvature tracks are required, 
with reconstructed transverse momenta 
$p_{T1}, p_{T2} > 2$
GeV/$c$, the scalar sum $p_{T1}+p_{T2}>$ 5.5 GeV/$c$, and a 
transverse opening-angle 
$\Delta\phi < $~135$^{\circ}$.
At Level~2, the Silicon Vertex Trigger (SVT) \cite{svt} 
combines XFT tracks with SVX~II hits
to measure the impact parameter $d$ (distance of closest approach to the beam line)
 of each valid track.
The requirement of two tracks with 100~$\mu$m~$< 
d < 1.0$~mm reduces light-quark background by two orders of magnitude while 
preserving $\simeq 50\%$ of the signal. 
A tighter opening angle cut: $20^{\circ} < \Delta\phi < $~135$^{\circ}$
selects two-body $B$ decays from multi-body with $97\%$ efficiency and reduces 
background further.
Each track pair is then used to form a $B$ candidate, which is 
required to have an impact parameter relative to the beam axis 
$d_B<140\ \mu\rm{m}$ 
and to have travelled a transverse distance $L_{xy} >$~200~$\mu$m.  
At Level~3, a farm of computers confirms the selection with a full event reconstruction.
The overall acceptance of the trigger selection is $\simeq 2\% $ for $B$ mesons 
of $p_{T}>4$~GeV/$c$.  

In the offline analysis, combinatoric and light-quark backgrounds
are effectively rejected by requiring the $B$ candidate to be isolated. 
The isolation cut ($I > 0.5$)~\cite{Isolation} has been 
chosen, together with tightened cuts on kinematic observables 
($L_{xy} > 300\ \mu$m, $d_B < 80\ \mu$m, and $d > 150\ \mu$m), by 
maximizing the quantity S/(S+B)$^{1/2}$ over all possible 
combinations of cuts.  
The background B is estimated from the data sidebands. 
The expected signal yield S is obtained from a detailed detector
simulation 
assuming the momentum distribution of $B$-mesons measured by 
CDF~\cite{B-PT}, and normalized to 
the yield observed after the trigger selection. The overall 
efficiency of the chosen offline selection is $\simeq 50\%$. 

\begin{figure}[htb]
\includegraphics[width=160pt]{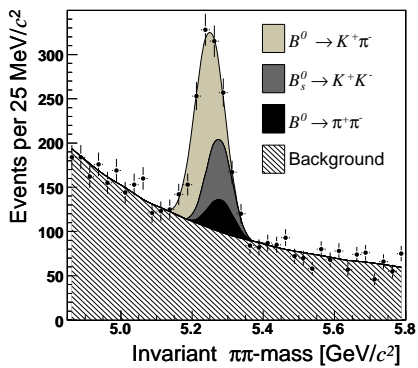}
\caption{\label{fig:b0hh}
Invariant mass distribution
of \Bhhgen\ candidates passing all selection requirements, using a
pion mass assumption for both decay products. 
Cumulative projections of the likelihood fit for each mode are 
overlaid.}
\end{figure}

No more than one $B$ meson
candidate per event survives the selection, and a mass is assigned to 
each, assuming the pion mass for both decay products. The mass 
distribution, shown in Fig.~\ref{fig:b0hh}, exhibits an obvious peak 
in the $B^{0}_{(s)}$ mass region. 
A binned fit of a Gaussian over an exponential background provides an
estimate of $893\pm 47$ signal events, with a width $\sigma = 
38\pm 2$~MeV/$c^{2}$, compared to an expected mass resolution $\sigma 
= 28$~MeV/$c^{2}$ for an individual \Bhhgen\ mode. This indicates the 
presence of at least two distinct final states.
Sizable signal contributions are expected from two known \Bd\ modes, \Bdpipi\ and
\BdKpi , and two as yet unobserved \Bs\ modes, \BsKK\ and \BsKpi .  
Figure~\ref{fig:b0hh} shows
that, as expected, the different modes are too closely spaced in
mass to be clearly resolved and appear instead as a single peak
somewhat broader than the mass resolution.  
In addition to mass resolution, we use kinematic information along 
with particle identification to extract the different contribution.
We incorporate all information in an unbinned likelihood fit,
to statistically determine the contribution of each mode,
and the CP asymmetry of the \BdKpi\ mode 
$A_{CP}=[N(\bar{B})-N(B)]/[N(\bar{B})+N(B)]$.

\begin{figure}[htb]
\includegraphics[height=4.0 cm]{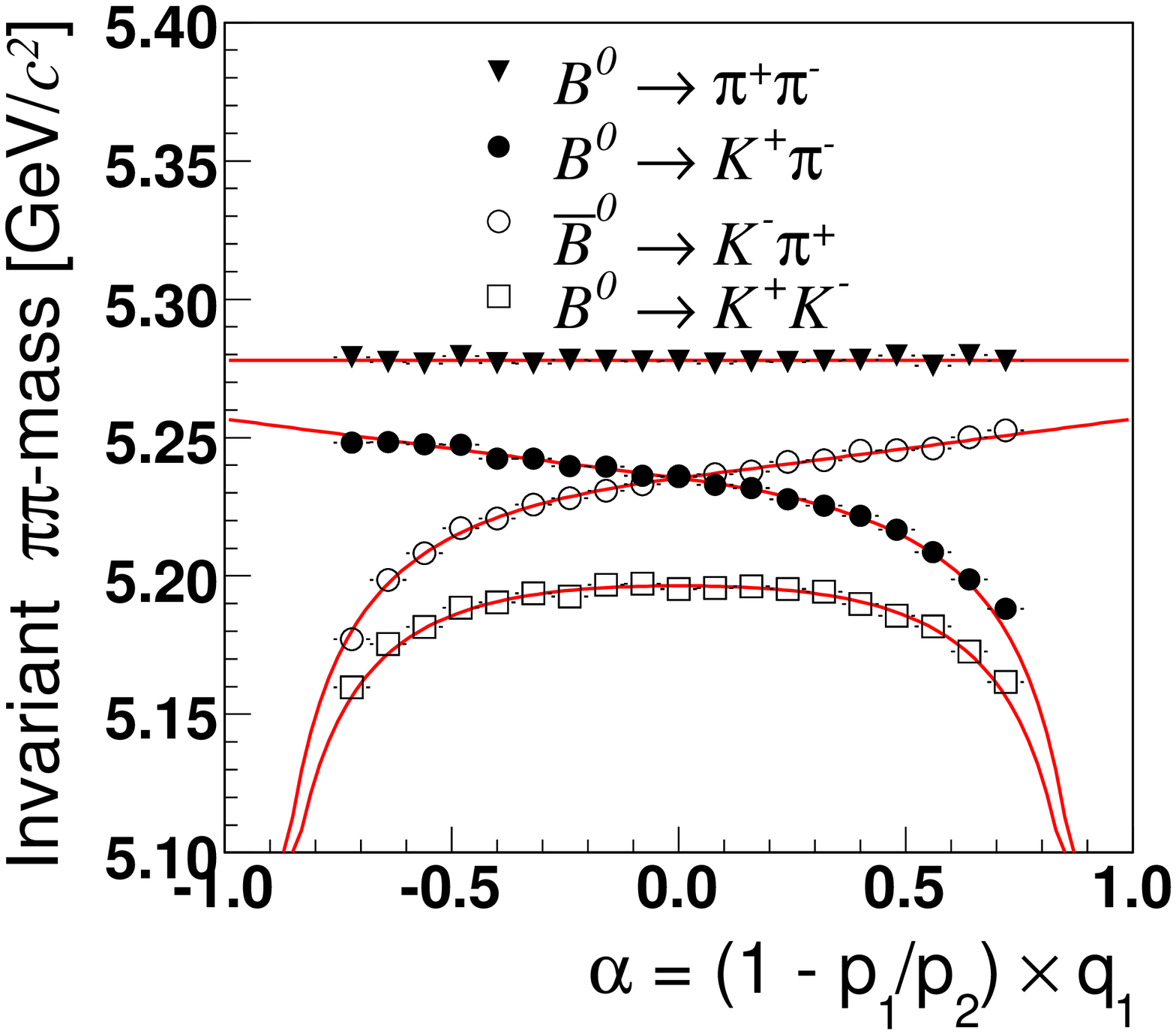}
\includegraphics[height=4.0 cm]{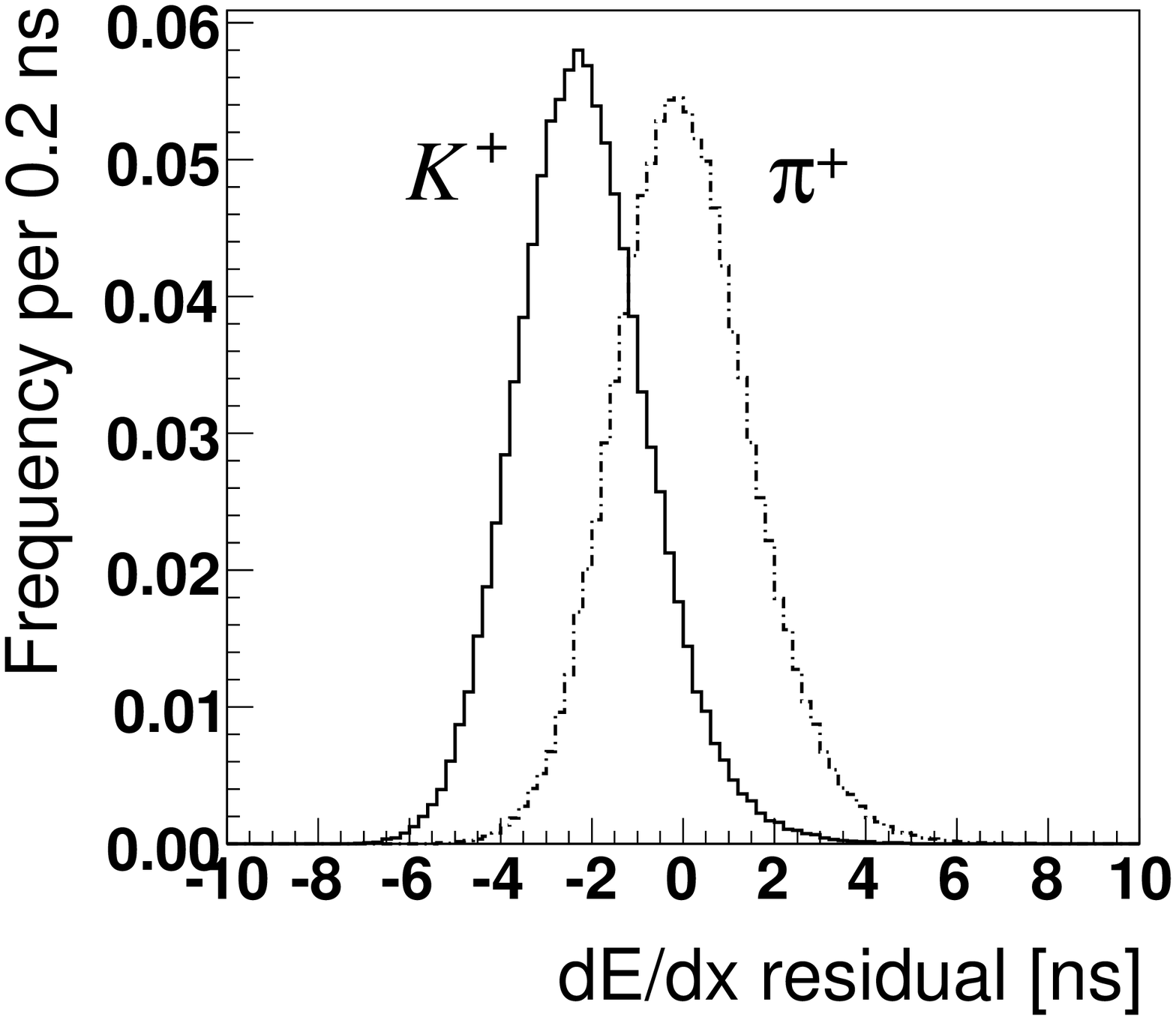}
\caption{(left) Average $M_{\pi\pi}$ versus $\alpha$ for simulated 
samples of 
\Bd\ events, where $K^{+}\pi^{-}$ and $K^{-}\pi^{+}$ are treated separately. 
The solid curves are the corresponding first-order expressions 
from Eq.~(\ref{eq:Mhh}). The corresponding plots for the \Bs\ are 
similar, but shifted for the mass difference. (right) Distribution of
\dedx\ (mean COT pulse-width) around the average pion response, 
for calibration samples of kaons and pions (see text).
\label{fig:Mpipivsalpha}}

\end{figure}

For the kinematic portion, we use two loosely correlated 
observables to summarize the 
information carried by all possible values of invariant mass of the 
candidate $B$, 
resulting from different mass assignments to the two outgoing 
particles~\cite{punzibias}. They are the mass
$M_{\pi\pi}$ calculated with the pion mass assignment to both 
particles,
and the signed momentum imbalance 
$\alpha = (1-p_1/p_2) q_{1}$, where $p_1$ ($p_2$) is the  
lower (higher) of the particle momenta, and $q_1$ is the sign of the charge of the 
particle of momentum $p_{1}$. 
Using these two variables, the mass of any particular mode can be
expressed, in the relativistic limit, as:

{\footnotesize
\begin{equation}\label{eq:Mhh}
M^{2}_{m_{1} m_{2}} =  M^{2}_{\pi \pi} - 
( 2-|\alpha| ) ( m^{2}_{2} - m_{\pi}^{2} )
- [1-(|\alpha|-1)^{-1}] ( m^{2}_{1} - m_{\pi}^{2} )
\end{equation}
}
where $m_{1}$ ($m_{2}$) is the mass of the lower (higher) momentum 
particle (Fig.~\ref{fig:Mpipivsalpha}, left).
Particle identification (PID) information is provided by the measured \dedx\  
of the two tracks. In order to account for their dependence on particle momentum, 
we include in our fit the scalar sum $p_{\rm{tot}}=p_{1}+p_{2}$ as a fifth observable,
which in conjunction with $\alpha$ provides unique identification
of the momenta of both particles. 

 With the chosen observables, the likelihood 
 contribution of the $i^{\mathit{th}}$ event is written as:
\begin{equation}
    \mathcal{L}_i = (1-b)\sum_{j} f_j \mathcal{L}^{\mathrm{kin}}_j 
    \mathcal{L}^{\mathrm{PID}}_j +  
    b \mathcal{L}^{\mathrm{kin}}_{\mathrm{bck}} 
    \mathcal{L}^{\mathrm{PID}}_{\mathrm{bck}} \\
\end{equation}
    where the index `$\mathrm{bck}$' labels background-related quantities,
    the index $j$ runs over the eight distinguishable \Bhhgen\ modes 
    (Fig.~\ref{fig:Mpipivsalpha}), and 
    $f_j$ are their fractions, to be 
    determined by the fit together with the background fraction 
    $b$.
    The $\mathcal{L}^{\mathrm{kin}}_j$ is given by the product of the
    conditional probability density of $M_{\pi\pi}$ for 
    given $\alpha$ and the joint probability 
    distribution $P_{j}(\alpha,p_{\rm{tot}})$.
    The mass distribution is a Gaussian centered at  
    the value of $M_{\pi\pi}$ obtained from Eq.~(\ref{eq:Mhh})  
    by setting the appropriate particle masses for each decay mode $j$.
    The Gaussian width $\sigma_M = 28\pm 3$ MeV/$c^{2}$ was interpolated from 
    the observed widths of other two-body decays ($D^0\rightarrow 
    K^{-}\pi^{+}$, $J/\psi \rightarrow \mu^+\mu^-$, and $\Upsilon 
    \rightarrow \mu^+\mu^-$), and the \Bd\ and \Bs\ masses are set to 
    the values measured by CDF~\cite{CDFmasspaper} to cancel 
    the common systematic uncertainty.
    The background mass distribution is fitted to an exponential 
    function plus a constant. 
    The $P_{j}(\alpha,p_{\rm{tot}})$ is 
    parameterized for each mode $j$ by a product of 
    polynomial and exponential functions, 
    fitted to Monte Carlo samples 
    produced by a detailed detector simulation, while the 
    corresponding distribution for the background is obtained from the mass sidebands of data.

The \dedx\ response was calibrated over the tracking volume 
and time by means of a $97\%$-pure sample of $3\times 10^5$ 
$D^{*+}\rightarrow D^0\pi^+ 
 \rightarrow [K^-\pi^+]\pi^+$ decays, where the $D^0$ decay 
products are identified by the charge of the $D^{*+}$ pion~\cite{Diego}.  
The observed response (Fig.~\ref{fig:Mpipivsalpha}, right)
is well-modeled by the convolution of a single-particle response 
function with a common baseline fluctuation, causing a $10\%$ correlation 
between particles in the same event.  
Both effects are quasi-Gaussian with small tails, and have been 
accurately modeled in $\mathcal{L}^{\mathrm{PID}}$.
The separation between pions and kaons in the range 
$2<p_{T}<10$~GeV/$c$ is nearly constant at 1.4 
standard deviations, corresponding 
to a resolution 1.7 times worse than a ``perfect'' PID, when 
measuring the relative fractions of the two particles in any given sample.
The $\mathcal{L}^{\mathrm{PID}}_{\mathrm{bck}}$ term allows 
for independent pion, kaon, proton, and
electron components, which are free to 
vary independently in three mass regions (left, under, and right of 
the signal peak) to allow for possible variations due to the
contribution of partially-reconstructed $B$ hadrons in 
the lower-mass region. Muons are indistinguishable from pions with the 
available \dedx\ resolution.

The fit of the data sample returns the yields listed in Table~\ref{tab:yields}. 
The observed resolutions are compatible with expectations from fitting Monte Carlo samples
of the same size.  
Significant signals are seen for \Bdpipi , \BdKpi , and the 
previously unobserved \BsKK\ mode, while no evidence is obtained for 
\BsKpi , \Bspipi , or \BdKK .
As a check of our results, we 
performed an alternative fit based solely on kinematical information. 
Since the \Bdpipi\ mode is indistinguishable from \BsKK\ in absence 
of PID information, we constrain its rate to its world-average value~\cite{HFAG06}. 
This fit confirms the main results, returning 
a yield of $193 \pm 55$ \BsKK\ events.
%
To convert raw yields into relative branching fractions, 
we apply corrections for the different efficiencies of trigger and 
offline selection requirements for different decay modes; 
the relative efficiency corrections between modes do not exceed $19\%$.
Most corrections are determined from the detailed
detector simulation
, with the following exceptions 
which are measured using data: 
A momentum-averaged relative isolation efficiency between \Bs\ 
and \Bd\ of $1.07 \pm 0.11$ has been determined from fully-reconstructed samples of
\Bs $\rightarrow J/\psi\,\phi$, \Bs $\rightarrow D^{-}_{s}\pi^{+}$, 
\Bd $\rightarrow J/\psi\,K^{*0}$,
and \Bd $\rightarrow D^{-}\pi^{+}$.
The lower specific ionization of kaons with respect to pions in the 
COT is responsible for a 
$\simeq 5$\% lower efficiency to reconstruct a kaon by the XFT. 
This effect is measured in a sample of $D^{+}\rightarrow K^{-}\pi^{+}\pi^{+}$ 
decays triggered on two tracks, using the unbiased third track.
\begin{table*}[tb]
\centering
\caption{\label{tab:yields}\label{tab:BRs} Summary of results. The 
yields of the two annihilation modes (last two rows) were fixed to zero when 
fitting for the four main modes. Absolute branching fractions are 
normalized to the the world--average values 
${\mathcal B}(\mbox{\BdKpi}) = (18.9\pm 0.7) \times 10^{-6}$ and 
$f_{s}/f_{d}=0.26 \pm 0.039$~\cite{HFAG06}. 
The first quoted uncertainty is statistical, the second is systematic.}
\begin{tabular}{lr|rcl|ccr}
\hline\hline
	Mode			&      	   Yield~~~&	
	\multicolumn{3}{|c|}{Measured Quantity}		&	
	Derived \BR\ ($10^{-6}$) 			\\
\hline				  	       					           							        
	\bdkpi			&    	  542 $\pm$ 30	
	&	\multicolumn{3}{|c|}{$A_{\mathrm{CP}}$ = $-0.013$ 
	$\pm$ 0.078 $\pm$ 0.012}& 						   	\\
	\bdpipi          	&    	  121 $\pm$ 27	&	
	\fracpipisukpi	&=&   	0.21 $\pm$ 0.05 $\pm$ 0.03	&	3.9 $\pm$ 1.0 $\pm$ 0.6\\ 
	\bskk            	&    	  236 $\pm$ 32	&	
	\frackksukpi	&=&   	0.46 $\pm$ 0.08 $\pm$ 0.07	&       33 $\pm$ 6 $\pm$ 7\\ 
	\bskpi           	&    	    3 $\pm$ 25 &	
	\fraclimbskpi 	&$<$& 0.08 @~90\%~CL	        &	$< 5.6$ @~90\%~CL 	\\ 
	\hline
	\bspipi          	&    	 $-10$ $\pm$ 15	&	
	\fraclimbspipi	&$<$&   0.05 @~90\%~CL    	        &	$< 1.7$ @~90\%~CL	\\ 
	\bdkk            	&    	   10 $\pm$ 23	&	
	\fraclimbdkk  	&$<$&             0.10 @~90\%~CL	        &	$< 1.8$ @~90\%~CL	\\ 
\hline\hline
\end{tabular}
\end{table*}
The only correction needed by $A_{\mathrm{CP}}$ is a $(1.0 \pm 
0.25)\%$ shift due to the different probability for $K^{+}$ and 
$K^{-}$ to interact with the tracker material.
The accuracy of our control over instrumental charge asymmetries is confirmed by the 
smallness of the asymmetry ($<0.5\%$) measured in the $D^{0}\rightarrow 
K^{-}\pi^{+}$ mode~\cite{D0paper}.
The \BsKK\ and \Bspipi\ modes require a special treatment, since they 
contain a superposition of the flavor eigenstates of the \Bs . 
Their time evolution might differ from the flavor-specific 
modes if the width difference $\Delta\Gamma_{s}$ 
between the \Bs\ mass eigenstates is significant.  
The current result is derived under the assumption that both modes are
dominated by the short-lived \Bs\ component
, that $\Gamma_s=\Gamma_d$, and $\Delta\Gamma_s/\Gamma_s = 
0.12\pm 0.06$~\cite{Beneke:1998sy}. 
The latter uncertainty is included in estimating the overall systematic 
uncertainty. 

 %
The dominant contributions to the systematic uncertainty are: 
the statistical uncertainty on isolation efficiency (\Bs\ modes), 
possible charge asymmetry of background ($A_{\mathrm{CP}}$), and final 
state photon radiation (\Bdpipi). The latter is conservatively 
estimated with the full effect predicted by QED calculations~\cite{Cirigliano-Isidori}.
Smaller systematic uncertainties are assigned for: 
mass scale and resolution; 
\dedx\ response 
model
; trigger efficiencies; background shape and kinematics;  
$B$ meson masses, lifetimes, and differences in momenta, allowed to
vary by a factor $(m_{\bs}-m_{\bd})/m_{\bd}$ due to fragmentation 
effects~\cite{Mangano}.
 %

 The relative branching fractions obtained after applying all corrections are 
 listed in Table~\ref{tab:BRs}, where $f_{d}$ and $f_{s}$ indicate 
 the production fractions respectively of \Bd\ and \Bs\  
 from fragmentation of a $b$ quark in $\bar{p}p$ collisions. 
 Upper limits are quoted for modes in which no significant signal is
 observed~\cite{F-C}. We also list 
 absolute results obtained by normalizing our data to the world-average 
 of \BR(\BdKpi) and assuming for $f_{s}/f_{d}$ the world-average from 
 \ppbar\ and $e^{+}e^{-}$ experiments~\cite{HFAG06}.
 
 The rate of the newly observed mode \BsKK\ favors the higher value 
 $(36\pm 7)\times 10^{-6}$ predicted by 
 calculations based on QCD sum rules~\cite{Khodjamirian,Burasetal} 
 implying large U-spin breaking in this process, 
 although it is not statistically incompatible with the expectation 
 \BR(\BsKK) $ = $ \BR(\BdKpi) from the assumption of exact U-spin symmetry
 and negligible spectator contributions. 
 We also derive the ratio of U-spin--conjugate decays: 
 $(\mathit{f_d}/\mathit{f_s})\times\mathcal{B}(\bdpipi)/\mathcal{B}(\bskk)$ = 0.45 $\pm$ 
 0.13 $\pm$ 0.06, which can be related to the CP asymmetries 
 in the \Bdpipi\ mode and to the CKM angle $\gamma$~\cite{Fleischer}.
 Our results for the \Bd\ are in agreement with world average values: 
 \BR(\Bdpipi) $=(4.6\pm 0.4)\times 10^{-6}$ and 
 $A_{\mathrm{CP}}(\mbox{\BdKpi}) = -0.113\pm 0.020$~\cite{PDG2006}
 although our $A_{\mathrm{CP}}$ measurement is also compatible with 
 zero. 
 The limit set on \BsKpi\ indicates a 
 value at the lower end of current expectations~\cite{B-N,Bs-QCDF}.
 The limit for the annihilation mode \Bspipi\ is a large 
 improvement over the previous best limit~\cite{Bspipioldlimit}, 
 approaching the expectations from recent calculations~\cite{Bspipi,Bspipi-PQCD}.

 
 In summary, we have measured relative branching fractions of 
 $B^{0}_{(s)}$ mesons into pairs of charmless charged mesons. 
We find results in agreement with current world averages for \Bd\ 
modes and observe for the first time the \BsKK\ mode. 
We set upper limits on unobserved modes \BdKK, \BsKpi, and \Bspipi .

\begin{acknowledgments}
We thank the Fermilab staff and the technical staffs of the participating
institutions for their vital contributions.  This work was supported by the
U.S. Department of Energy and National Science Foundation; the Italian
Istituto Nazionale di Fisica Nucleare; the Ministry of Education, Culture,
Sports, Science and Technology of Japan; the Natural Sciences and
Engineering Research Council of Canada; the National Science Council of the
Republic of China; the Swiss National Science Foundation; the A.P. Sloan
Foundation; the Bundesministerium f\"ur Bildung und Forschung, Germany; the
Korean Science and Engineering Foundation and the Korean Research
Foundation; the Particle Physics and Astronomy Research Council and the
Royal Society, UK; the Russian Foundation for Basic Research; the
Comisi\'on Interministerial de Ciencia y Tecnolog\'{\i}a, Spain; the
European Community's Human Potential Programme under contract
HPRN-CT-2002-00292; and the Academy of Finland.

\end{acknowledgments}
 %
 %

\begin{thebibliography}{99}
\bibitem{CDF-coordinates} CDF II uses a cylindrical coordinate
system in which $\phi$ is the azimuthal angle, $r$ is the radius
from the nominal beam line, and $z$ points in the proton beam 
direction, with the origin at the center of the detector. The transverse
plane is the plane perpendicular to the $z$ axis.
\bibitem{Isolation} Isolation is defined 
as $I= p_{T}(B)/(p_{T}(B)+\sum_{i} p_{T i})$, 
where $p_{T}(B)$ is the transverse momentum of the $B$ candidate,
and the sum runs over all other
tracks within a cone of radius 1 in $\eta$-$\phi$ space around the $B$ flight-direction.
\end{thebibliography}
 %

\end{document}